\documentclass[conference,letterpaper]{IEEEtran}
\IEEEoverridecommandlockouts

\usepackage[T1]{fontenc}
\usepackage{times}
\usepackage{amsmath,amssymb,amsfonts}
\usepackage{bm}
\usepackage[dvips]{graphicx}
\DeclareGraphicsExtensions{.eps}
\usepackage{xcolor}
\usepackage{booktabs}
\usepackage{multirow}
\usepackage{array}
\usepackage{url}
\usepackage{balance}
\usepackage{cite}
\usepackage{microtype}
\usepackage[caption=false,font=footnotesize]{subfig}
\usepackage{tikz}
\usetikzlibrary{arrows.meta, positioning, fit, calc, backgrounds}
\graphicspath{{figures/}}

\begin{document}
\title{Measuring the Value of Communication in Multi-Agent Tracking Under Finite-Rate Belief Sharing}

\author{
\IEEEauthorblockN{Danial Dehghani, Maximilian L\"ubke, Zahra Seifaei and Norman Franchi}
\IEEEauthorblockA{Institute for Smart Electronics and Systems (LITES)\\
Friedrich-Alexander-Universit\"at Erlangen-N\"urnberg\\
Email: \{danial.dehghani, maximilian.luebke, zahra.seifaei, norman.franchi\}@fau.de}
}

\maketitle
\begin{abstract}
This paper investigates the value of communication (VoC) in cooperative multi-agent tracking under rate-constrained information exchange. Although prior work has studied distributed tracking, quantized communication, and robust fusion, it remains unclear at what communication rate most of the achievable tracking gain is obtained, beyond which additional communication resources yield only marginal benefit. We consider stationary agents that observe a moving target through heterogeneous range and bearing measurements, perform local extended Kalman filtering (EKF), quantize their posterior state means, and fuse exchanged information via covariance intersection (CI). Performance is evaluated versus communication rate relative to no-communication baselines, a centralized oracle, and a CI-perfect reference. The results show that communication is not uniformly beneficial: very low rates can degrade performance due to coarse quantization, whereas moderate rates recover most of the achievable gain. In the considered scenario, increasing the rate from 4 to 12 \(\text{bit/s}\) per agent reduces the root-mean-square error (RMSE) from 13.93\,m to 6.35\,m, while a further increase to 32 \(\text{bit/s}\) per agent improves it only marginally to 5.93\,m. This reveals a clear knee region at moderate rates, beyond which the remaining loss is mainly due to the conservatism of the fusion architecture. Overall, the proposed framework characterizes low-rate degradation, communication-efficient operation, and diminishing returns, while highlighting the joint role of rate and sensing geometry in determining communication benefit.
\end{abstract}

\begin{IEEEkeywords}
value of communication, multi-agent tracking, extended Kalman filter, covariance intersection, distributed state estimation, rate-constrained communication
\end{IEEEkeywords}
\section{Introduction}
Multi-agent systems (MAS) such as connected vehicles, cooperative robots, and distributed sensor networks can benefit from communication and information exchange to improve task performance~\cite{gielis2022critical,wang2023survey}. However, communication consumes bandwidth, introduces latency, and incurs energy and privacy costs, which often make it a system bottleneck~\cite{hu2024overview}. Communication is therefore not inherently valuable; its benefit must be evaluated relative to the resources it consumes. This motivates the central question considered in this paper: \emph{how much inter-agent communication is justified by the resulting performance improvement, relative to its communication cost?}

This question is especially important in cooperative tracking, where each agent has only local and incomplete information about the target state. In such settings, communication can improve global estimation performance, but its utility depends strongly on how informative the exchanged messages are, how restrictive the available rate constraint is, and how the received information is fused.

Prior work most relevant here falls into three threads. Distributed estimation over sensor
networks combines local estimates for global accuracy~\cite{he2020distributed,souza2016target},
but fixes the protocol rather than the rate itself. Rate limited and event triggered schemes
bound how much is sent~\cite{battistelli2018distributed,zhu2015distributed}, but emphasize
stability and triggering over the gain a given rate buys. Communication aware and task oriented
exchange learns what to transmit~\cite{jiang2018learning,das2019tarmac}, yet evaluates whole
policies without isolating the rate accuracy tradeoff. It thus remains unclear how communication
benefit evolves with rate in cooperative tracking, and where added rate ceases to be justified.

A meaningful value of communication (VoC) analysis also requires a consistent fusion
model. Because repeated exchange of local estimates induces unknown statistical
dependencies among agents~\cite{bakr2017distributed,li2023overview}, ignoring them can
overstate the apparent VoC; we therefore adopt the conservative covariance intersection
(CI) rule~\cite{ajgl2018ci}. Unlike the threads above, we keep the estimator, quantizer,
and fusion rule fixed and resolve tracking distortion as an explicit function of
communication rate, separating loss due to limited rate from that due to conservative
fusion.

In this paper, we study cooperative multi-agent tracking under rate-constrained communication and analyze how tracking performance evolves with communication rate. To this end, we construct VoC curves that relate communication rate to tracking distortion and reveal where communication is harmful, beneficial, or subject to diminishing returns.

We evaluate the proposed framework against three references: no communication, a centralized oracle with access to all raw agent measurements, and a CI-perfect case with unquantized posterior exchange under the same CI fusion rule. Together with geometry-aware diagnostics, these references enable the interpretation of communication-limited and fusion-limited regimes.

The main contributions of this work are:
\begin{itemize}
    \item a task-oriented VoC metric and evaluation framework for cooperative tracking under rate-constrained communication;
    \item a reference-based methodology using no-communication, centralized-oracle, and CI-perfect baselines to distinguish communication-limited from fusion-limited behavior; and
    \item a geometry-aware analysis of when communication is beneficial and when additional rate is no longer justified.
\end{itemize}
The considered setup is intentionally controlled to isolate the effect of finite-rate belief exchange; while exact operating points are scenario-dependent, the resulting VoC trends provide broader design insight for cooperative tracking.

The remainder of this paper is organized as follows. Section~II presents the system model. Section~III introduces the VoC metrics, reference baselines, and evaluation diagnostics. Section~IV presents the simulation setup and numerical results. Finally, Section~V concludes the paper.
\section{System Model}

Figure~\ref{fig:system_overview} summarizes the workflow considered in this paper. The shared system model consists of target dynamics in Fig.~\ref{fig:system_overview}(a) and sensing and measurements in Fig.~\ref{fig:system_overview}(b), where a moving target is observed by multiple stationary agents through noisy measurements. Based on this common sensing stage, three estimation branches are considered: no communication in Fig.~\ref{fig:system_overview}(c), where each agent performs local state estimation using only its own measurements; oracle in Fig.~\ref{fig:system_overview}(d), where all raw measurements are available to a centralized estimator; and finite-rate communication in Fig.~\ref{fig:system_overview}(e), where each agent computes a local posterior estimate, quantizes its posterior state mean, broadcasts the resulting rate-limited message, and then applies CI to the common set of quantized posterior means, including its own. This symmetric fusion architecture ensures that all CI inputs have the same rate-limited representation, which allows the effect of communication rate to be analyzed in a controlled and interpretable manner. In addition to the three main branches in Fig.~\ref{fig:system_overview}(c)--(e), we also consider a CI-perfect reference obtained by removing quantization from the finite-rate branch in Fig.~\ref{fig:system_overview}(e) under the same CI fusion rule.


\definecolor{agent1col}{RGB}{55,110,120}
\definecolor{agent2col}{RGB}{88,102,160}
\definecolor{agentNcol}{RGB}{118,136,84}

\begin{figure}[t]
\centering
\resizebox{\columnwidth}{!}{%
\begin{tikzpicture}[
    font=\rmfamily\footnotesize,
    every node/.style={font=\rmfamily\footnotesize},
    block/.style={
        rectangle, draw=black!70, fill=white,
        minimum height=2.55em, minimum width=6.8em,
        align=center, font=\rmfamily\footnotesize,
        rounded corners=2pt, line width=0.5pt,
        inner sep=3.5pt
    },
    wideblock/.style={
        rectangle, draw=black!70, fill=white,
        minimum height=4.2em, minimum width=4.55cm,
        align=center, font=\rmfamily\footnotesize,
        rounded corners=2pt, line width=0.5pt,
        inner sep=3.5pt
    },
    caseinner/.style={
        rectangle, draw=black!60, fill=white,
        minimum height=2.45em, minimum width=3.85cm,
        align=center, font=\rmfamily\footnotesize,
        rounded corners=1.5pt, line width=0.45pt,
        inner sep=2.7pt
    },
    panelinner/.style={
        rectangle, draw=black!60, fill=white,
        minimum height=3.0em, minimum width=3.85cm,
        align=center, font=\rmfamily\footnotesize,
        rounded corners=1.5pt, line width=0.45pt,
        inner sep=2.7pt
    },
    agentinner/.style={
        rectangle, draw=black!60, fill=white,
        minimum height=1.95em, minimum width=0.78cm,
        align=center, font=\rmfamily\scriptsize,
        rounded corners=1.2pt, line width=0.4pt,
        inner sep=1.1pt
    },
    agentone/.style={
        agentinner,
        draw=agent1col,
        text=agent1col
    },
    agenttwo/.style={
        agentinner,
        draw=agent2col,
        text=agent2col
    },
    agentNstyle/.style={
        agentinner,
        draw=agentNcol,
        text=agentNcol
    },
    dotsinner/.style={
        rectangle, draw=none, fill=none,
        minimum height=1.95em, minimum width=0.26cm,
        align=center, font=\rmfamily\footnotesize,
        inner sep=0.4pt
    },
    ghost/.style={
        rectangle, draw=none, fill=none,
        minimum height=2.45em, minimum width=3.85cm,
        inner sep=0pt
    },
    envbox/.style={
        rectangle, draw=black!45, fill=gray!4,
        fill opacity=0.35, text opacity=1,
        rounded corners=3pt, line width=0.55pt
    },
    ncenvbox/.style={
        rectangle, draw=black!50, fill=black!3,
        fill opacity=0.45, text opacity=1,
        rounded corners=3pt, line width=1.1pt
    },
    finenvbox/.style={
        rectangle, draw=agent1col!55!black, fill=agent1col!8,
        fill opacity=0.45, text opacity=1,
        rounded corners=3pt, line width=1.1pt
    },
    orenvbox/.style={
        rectangle, draw=agent2col!55!black, fill=agent2col!8,
        fill opacity=0.45, text opacity=1,
        rounded corners=3pt, line width=1.1pt
    },
    mainarr/.style={-Stealth, line width=0.55pt, black!75},
    arrone/.style={-Stealth, line width=0.55pt, draw=agent1col},
    arrtwo/.style={-Stealth, line width=0.55pt, draw=agent2col},
    arrN/.style={-Stealth, line width=0.55pt, draw=agentNcol},
    lbl/.style={font=\rmfamily\scriptsize, text=black!60},
    nclbl/.style={font=\rmfamily\scriptsize, text=black!65},
    finlbl/.style={font=\rmfamily\scriptsize, text=agent1col!65!black},
    orlbl/.style={font=\rmfamily\scriptsize, text=agent2col!65!black},
    boxmark/.style={
        font=\rmfamily\scriptsize\bfseries,
        text=black!70,
        fill=white,
        draw=none,
        inner sep=1.2pt,
        outer sep=0pt
    }
]

\def\xNC{1.8}
\def\yNC{3.35}

\def\xDyn{2.9}
\def\xMeas{7.4}
\def\ySys{.5}

\def\xOR{11.7}
\def\yOR{3.35}

\def\xFIN{7.4}
\def\yFinTop{-1.75}
\def\yFinMid{-3.75}
\def\yFinBot{-5.75}

\node[block] (dyn) at (\xDyn,\ySys) {Target\\Dynamics};

\node[wideblock, minimum height=4.35em] (meas) at (\xMeas,\ySys) {};
\node[font=\rmfamily\footnotesize] at ([yshift=0.48cm]meas.center) {Sensors and Measurements};

\node[agentone]    (m1)    at ([xshift=-1.08cm,yshift=-0.24cm]meas.center) {Agent\\1};
\node[agenttwo]    (m2)    at ([xshift=-0.30cm,yshift=-0.24cm]meas.center) {Agent\\2};
\node[dotsinner]   (mdots) at ([xshift=0.35cm,yshift=-0.24cm]meas.center) {$\cdots$};
\node[agentNstyle] (mN)    at ([xshift=1.00cm,yshift=-0.24cm]meas.center) {Agent\\$N$};

\draw[mainarr] (dyn.east) -- (meas.west);

\node[
    envbox,
    fit=(dyn)(meas),
    inner sep=9pt,
    label={[lbl, anchor=north west]north west:System Model}
] (sharedbox) {};

\node[boxmark, anchor=south west] at ([xshift=1pt,yshift=1pt]dyn.north west) {(a)};
\node[boxmark, anchor=south west] at ([xshift=1pt,yshift=1pt]meas.north west) {(b)};

\node[panelinner, minimum height=3.55em] (nclocal) at (\xNC,\yNC) {};
\node[agentone]    (nc1)    at ([xshift=-1.08cm,yshift=0.16cm]nclocal.center) {Agent\\1};
\node[agenttwo]    (nc2)    at ([xshift=-0.28cm,yshift=0.16cm]nclocal.center) {Agent\\2};
\node[dotsinner]   (ncdots) at ([xshift=0.35cm,yshift=0.16cm]nclocal.center) {$\cdots$};
\node[agentNstyle] (ncN)    at ([xshift=1.00cm,yshift=0.16cm]nclocal.center) {Agent\\$N$};
\node[font=\rmfamily\footnotesize] at ([yshift=-0.36cm]nclocal.center) {Local State 
Estimation};

\node[ghost] (nctop) at (\xNC,\yNC+0.82) {};
\node[ghost] (ncbot) at (\xNC,\yNC-0.82) {};

\node[
    ncenvbox,
    fit=(nctop)(nclocal)(ncbot),
    inner sep=9pt,
    label={[nclbl, anchor=north west]north west:No Communication}
] (ncbox) {};

\node[boxmark, anchor=south west] at ([xshift=1pt,yshift=1pt]ncbox.north west) {(c)};

\node[caseinner] (orest) at (\xOR,\yOR) {Centralized State\\Estimation};

\node[ghost] (ortop) at (\xOR,\yOR+0.82) {};
\node[ghost] (orbot) at (\xOR,\yOR-0.82) {};

\node[
    orenvbox,
    fit=(ortop)(orest)(orbot),
    inner sep=9pt,
    label={[orlbl, anchor=north west]north west:Oracle}
] (orbox) {};

\node[boxmark, anchor=south west] at ([xshift=1pt,yshift=1pt]orbox.north west) {(d)};

\node[panelinner, minimum height=3.65em] (finest) at (\xFIN,\yFinTop) {};
\node[agentone]    (fe1)    at ([xshift=-1.08cm,yshift=0.10cm]finest.center) {Agent\\1};
\node[agenttwo]    (fe2)    at ([xshift=-0.28cm,yshift=0.10cm]finest.center) {Agent\\2};
\node[dotsinner]   (fedots) at ([xshift=0.35cm,yshift=0.10cm]finest.center) {$\cdots$};
\node[agentNstyle] (feN)    at ([xshift=1.00cm,yshift=0.10cm]finest.center) {Agent\\$N$};
\node[font=\rmfamily\footnotesize] at ([yshift=-0.42cm]finest.center) {Local State Estimation};

\node[panelinner, minimum height=3.65em] (finquant) at (\xFIN,\yFinMid) {};
\node[agentone]    (fq1)    at ([xshift=-1.08cm,yshift=0.10cm]finquant.center) {Agent\\1};
\node[agenttwo]    (fq2)    at ([xshift=-0.28cm,yshift=0.10cm]finquant.center) {Agent\\2};
\node[dotsinner]   (fqdots) at ([xshift=0.35cm,yshift=0.10cm]finquant.center) {$\cdots$};
\node[agentNstyle] (fqN)    at ([xshift=1.00cm,yshift=0.10cm]finquant.center) {Agent\\$N$};
\node[font=\rmfamily\footnotesize] at ([yshift=-0.44cm]finquant.center) {Quantization \& Communication};

\node[caseinner] (finfuse) at (\xFIN,\yFinBot) {Data Fusion};

\node[
    finenvbox,
    fit=(finest)(finquant)(finfuse),
    inner sep=9pt,
    label={[finlbl, anchor=north west]north west:Finite-rate Communication}
] (finbox) {};

\node[boxmark, anchor=south west] at ([xshift=1pt,yshift=1pt]finbox.north west) {(e)};


\draw[arrone] (m1.north) -- (nc1.south);
\draw[arrtwo] (m2.north) -- (nc2.south);
\draw[arrN]   (mN.north) -- (ncN.south);

\draw[arrone] (m1.north) -- ([xshift=-0.90cm]orest.south);
\draw[arrtwo] (m2.north) -- (orest.south);
\draw[arrN]   (mN.north) -- ([xshift=0.90cm]orest.south);

\draw[arrone] (m1.south) -- (fe1.north);
\draw[arrtwo] (m2.south) -- (fe2.north);
\draw[arrN]   (mN.south) -- (feN.north);

\draw[arrone] (fe1.south) -- (fq1.north);
\draw[arrtwo] (fe2.south) -- (fq2.north);
\draw[arrN]   (feN.south) -- (fqN.north);
\draw[mainarr] (finquant.south) -- (finfuse.north);

\end{tikzpicture}%
}
\caption{System overview. The shared system model consists of (a) target dynamics and (b) sensors and measurements. The three evaluation cases are (c) no communication, (d) oracle, and (e) finite-rate communication with local estimation, quantization, and fusion.}
\label{fig:system_overview}
\end{figure}
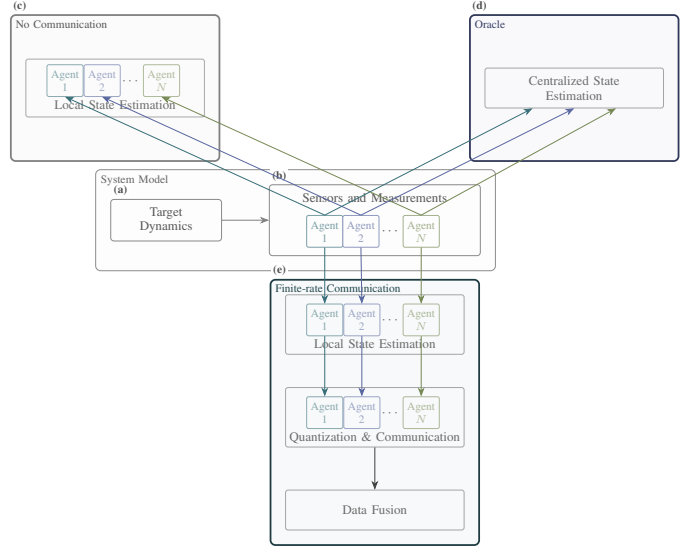
\subsection{Target Dynamics}
\label{subsec:target_dynamics}

The target is modeled in discrete time by a two-dimensional constant-velocity (CV) process driven by white Gaussian acceleration noise. Its state at time step $t$ is
\begin{equation}
\bm{x}_t =
\begin{bmatrix}
p_x(t) & p_y(t) & v_x(t) & v_y(t)
\end{bmatrix}^{\top}
\in \mathbb{R}^4,
\label{eq:state_vector}
\end{equation}
where $p_x(t)$ and $p_y(t)$ denote the target position in the $x$- and $y$-directions, and $v_x(t)$ and $v_y(t)$ denote the corresponding velocity components. The initial state $\bm{x}_0$ is specified in the simulation setup.

The state evolves according to
\begin{equation}
\bm{x}_{t+1} = \bm{A}\bm{x}_t + \bm{w}_t,
\qquad
\bm{w}_t \sim \mathcal{N}(\bm{0}, \bm{Q}),
\label{eq:target_dyn}
\end{equation}
where $\Delta t > 0$ denotes the sampling interval, $\bm{A}$ is the state transition matrix, and $\bm{w}_t$ is the process noise. Under the CV model,
\begin{equation}
\bm{A} =
\begin{bmatrix}
1 & 0 & \Delta t & 0\\
0 & 1 & 0 & \Delta t\\
0 & 0 & 1 & 0\\
0 & 0 & 0 & 1
\end{bmatrix}.
\label{eq:state_transition}
\end{equation}

Assuming identical and independent acceleration-noise processes in the $x$- and $y$-directions, each with standard deviation $\sigma_a$, the corresponding process-noise covariance is
\begin{equation}
\bm{Q}
=
\sigma_a^2
\begin{bmatrix}
\Delta t^4/4 & 0 & \Delta t^3/2 & 0\\
0 & \Delta t^4/4 & 0 & \Delta t^3/2\\
\Delta t^3/2 & 0 & \Delta t^2 & 0\\
0 & \Delta t^3/2 & 0 & \Delta t^2
\end{bmatrix}.
\label{eq:process_cov}
\end{equation}

For the simulated scenario, the target trajectory is generated recursively from $\bm{x}_0$ by sampling $\bm{w}_t$ according to \eqref{eq:target_dyn}. Larger values of $\sigma_a$ correspond to stronger deviations from ideal CV motion and, therefore, to less predictable target trajectories.
\subsection{Sensors and Measurements}
\label{subsec:sensors_measurements}

We consider $N$ stationary sensing agents at known positions
\begin{equation}
\bm{s}_i =
\begin{bmatrix}
s_{i,x} & s_{i,y}
\end{bmatrix}^{\top}
\in \mathbb{R}^2,
\qquad i \in \{1,\dots,N\}.
\label{eq:agent_position}
\end{equation}

At time step $t$, agent $i$ acquires a nonlinear range and bearing measurement of the target,
\begin{equation}
\bm{z}_{i,t} =
\begin{bmatrix}
r_{i,t} & \theta_{i,t}
\end{bmatrix}^{\top}
\in \mathbb{R}^2,
\label{eq:measurement_vector}
\end{equation}
where $r_{i,t}$ and $\theta_{i,t}$ denote the measured range and bearing, respectively. The measurement model is
\begin{equation}
\bm{z}_{i,t} = \bm{h}(\bm{x}_t,\bm{s}_i) + \bm{n}_{i,t},
\label{eq:meas_model}
\end{equation}
where $\bm{h}(\bm{x}_t,\bm{s}_i)$ denotes the corresponding noiseless range and bearing observation:
\begin{equation}
\bm{h}(\bm{x}_t,\bm{s}_i)
=
\begin{bmatrix}
\sqrt{(p_x(t)-s_{i,x})^2 + (p_y(t)-s_{i,y})^2}\\[0.3em]
\operatorname{atan2}\!\bigl(p_y(t)-s_{i,y},\, p_x(t)-s_{i,x}\bigr)
\end{bmatrix}.
\label{eq:measurement_function}
\end{equation}
Here, $\operatorname{atan2}(y,x)$ denotes the four-quadrant inverse tangent, i.e., the angle of the vector $(x,y)$ measured from the positive $x$-axis.

Thus, the first component gives the target range from agent $i$, while the second gives the corresponding bearing angle. Both depend only on the target position and not directly on the velocity components. The measurement noise is modeled as
\begin{equation}
\bm{n}_{i,t} \sim \mathcal{N}(\bm{0}, \bm{R}_i),
\qquad
\bm{R}_i = \operatorname{diag}\!\bigl(\sigma_{r,i}^2, \sigma_{\theta,i}^2\bigr),
\label{eq:measurement_noise}
\end{equation}
where $\sigma_{r,i}$ and $\sigma_{\theta,i}$ denote the range and bearing noise standard deviations of agent $i$, respectively. This formulation allows heterogeneous sensing quality across agents.
\subsection{Local State Estimation}
\label{subsec:local_state_estimation}

Each agent estimates the target state locally by first predicting the state using the motion model and then updating this prediction using its own measurement. To this end, agent $i$ maintains a local Gaussian belief
\begin{equation}
\bm{x}_t \mid \mathcal{H}^{\mathrm{loc}}_{i,t} \sim \mathcal{N}\!\left(\bm{\mu}_{i,t|t}, \bm{P}_{i,t|t}\right),
\label{eq:local_belief}
\end{equation}
where $\mathcal{H}^{\mathrm{loc}}_{i,t}$ denotes the local information available at agent $i$ up to time $t$, prior to communication and fusion, $\bm{\mu}_{i,t|t}$ is the local posterior mean, and $\bm{P}_{i,t|t}$ is the corresponding posterior covariance. All agents are initialized from the common prior
\begin{equation}
\bm{x}_0 \sim \mathcal{N}(\bm{\mu}_0, \bm{P}_0).
\label{eq:common_prior}
\end{equation}

Using the dynamics in \eqref{eq:target_dyn}, agent $i$ first computes the predicted mean and covariance
\begin{subequations}
\begin{align}
\bm{\mu}_{i,t|t-1} &= \bm{A}\bm{\mu}_{i,t-1|t-1}, \label{eq:ekf_pred_mean}\\
\bm{P}_{i,t|t-1} &= \bm{A}\bm{P}_{i,t-1|t-1}\bm{A}^{\top} + \bm{Q}. \label{eq:ekf_pred_cov}
\end{align}
\label{eq:ekf_prediction}
\end{subequations}

To account for the nonlinearity of \eqref{eq:meas_model}, the local measurement update is carried out using an extended Kalman filtering (EKF). The predicted measurement is
\begin{equation}
\hat{\bm{z}}_{i,t} = \bm{h}(\bm{\mu}_{i,t|t-1}, \bm{s}_i),
\label{eq:ekf_pred_meas}
\end{equation}
and the corresponding measurement Jacobian is
\begin{equation}
\bm{H}_{i,t}
=
\left.\frac{\partial \bm{h}(\bm{x}, \bm{s}_i)}{\partial \bm{x}}\right|_{\bm{x}=\bm{\mu}_{i,t|t-1}}
=
\begin{bmatrix}
\dfrac{\Delta x_{i,t}}{\rho_{i,t}} & \dfrac{\Delta y_{i,t}}{\rho_{i,t}} & 0 & 0 \\[0.8em]
-\dfrac{\Delta y_{i,t}}{\rho_{i,t}^{2}} & \dfrac{\Delta x_{i,t}}{\rho_{i,t}^{2}} & 0 & 0
\end{bmatrix},
\label{eq:ekf_jacobian}
\end{equation}
where
\begin{equation}
\Delta x_{i,t} = \mu_{i,t|t-1}^{(p_x)} - s_{i,x},
\qquad
\Delta y_{i,t} = \mu_{i,t|t-1}^{(p_y)} - s_{i,y},
\label{eq:delta_xy}
\end{equation}
and
\begin{equation}
\rho_{i,t} = \sqrt{\Delta x_{i,t}^{2} + \Delta y_{i,t}^{2}}.
\label{eq:rho_def}
\end{equation}

The innovation is
\begin{equation}
\tilde{\bm{z}}_{i,t} = \bm{z}_{i,t} - \hat{\bm{z}}_{i,t},
\label{eq:innovation}
\end{equation}
where the bearing component is wrapped to $(-\pi,\pi]$. The EKF innovation covariance $\bm{S}_{i,t}$, Kalman gain $\bm{K}_{i,t}$, posterior mean, and posterior covariance are then given by
\begin{subequations}
\begin{align}
\bm{S}_{i,t} &= \bm{H}_{i,t}\bm{P}_{i,t|t-1}\bm{H}_{i,t}^{\top} + \bm{R}_i, \label{eq:ekf_innov_cov}\\
\bm{K}_{i,t} &= \bm{P}_{i,t|t-1}\bm{H}_{i,t}^{\top}\bm{S}_{i,t}^{-1}, \label{eq:ekf_gain}\\
\bm{\mu}_{i,t|t} &= \bm{\mu}_{i,t|t-1} + \bm{K}_{i,t}\tilde{\bm{z}}_{i,t}, \label{eq:ekf_upd_mean}\\
\bm{P}_{i,t|t} &= \left(\bm{I} - \bm{K}_{i,t}\bm{H}_{i,t}\right)\bm{P}_{i,t|t-1}, \label{eq:ekf_upd_cov}
\end{align}
\label{eq:ekf_update}
\end{subequations}
where $\bm{I}$ denotes the identity matrix.

Thus, each agent produces a local posterior pair $(\bm{\mu}_{i,t|t}, \bm{P}_{i,t|t})$ from its own measurements and the common motion model.
\subsection{Quantization and Communication}
\label{subsec:comm_quant}

After local estimation at time $t$, agent $i$ constructs a rate-limited communication message from its posterior pair $(\bm{\mu}_{i,t|t}, \bm{P}_{i,t|t})$. Let $d_x$ denote the dimension of the state vector. To obtain a rate-constrained representation, the posterior mean correction is quantized, and the induced uncertainty is accounted for through covariance inflation. Although the quantization is applied to the correction term, the communicated state quantity remains a quantized posterior mean.

Specifically, the posterior mean correction relative to the local prediction is
\begin{equation}
\bm{\delta}_{i,t} = \bm{\mu}_{i,t|t} - \bm{\mu}_{i,t|t-1}.
\label{eq:mean_correction}
\end{equation}
For each state component $j \in \{1,\dots,d_x\}$, a symmetric clipping interval is constructed from the predicted covariance:
\begin{equation}
\ell_{i,t}^{(j)} = -k_{\sigma}\sqrt{[\bm{P}_{i,t|t-1}]_{jj}},
\qquad
u_{i,t}^{(j)} = +k_{\sigma}\sqrt{[\bm{P}_{i,t|t-1}]_{jj}},
\label{eq:clip_bounds}
\end{equation}
where $k_{\sigma} > 0$ is a design parameter. The interval $[\ell_{i,t}^{(j)}, u_{i,t}^{(j)}]$ is then uniformly quantized with $L = 2^b$ levels, where $b$ denotes the number of bits used per state component. The corresponding quantization step size is
\begin{equation}
\Delta_{i,t}^{(j)} = \frac{u_{i,t}^{(j)} - \ell_{i,t}^{(j)}}{2^b}.
\label{eq:quant_step}
\end{equation}
Using a uniform midrise quantizer, the reconstructed correction is denoted by $\bm{\delta}_{i,t}^{\,q}$, which yields the transmitted mean
\begin{equation}
\bm{\mu}_{i,t|t}^{\,q} = \bm{\mu}_{i,t|t-1} + \bm{\delta}_{i,t}^{\,q}.
\label{eq:quantized_mean}
\end{equation}

The quantization error is approximated as independent across state components and uniformly distributed within each quantization bin. This gives the covariance approximation
\begin{equation}
\bm{\Sigma}_{q,i,t}
=
\operatorname{diag}\!\left(
\frac{(\Delta_{i,t}^{(1)})^2}{12},
\dots,
\frac{(\Delta_{i,t}^{(d_x)})^2}{12}
\right).
\label{eq:quant_cov}
\end{equation}
The covariance attached to the communicated message is then given by
\begin{equation}
\tilde{\bm{P}}_{i,t|t} = \bm{P}_{i,t|t} + \bm{\Sigma}_{q,i,t}.
\label{eq:inflated_cov}
\end{equation}

Hence, the finite-rate communicated message is
\begin{subequations}
\begin{align}
\mathcal{M}_{i,t} &= \bigl(\bm{\mu}_{i,t|t}^{\,q},\, \tilde{\bm{P}}_{i,t|t}\bigr), \label{eq:msg_quantized}\\
\mathcal{M}_{i,t}^{\mathrm{perf}} &= \bigl(\bm{\mu}_{i,t|t},\, \bm{P}_{i,t|t}\bigr), \label{eq:msg_perfect}
\end{align}
\label{eq:message_defs}
\end{subequations}
where \eqref{eq:msg_perfect} defines the CI-perfect reference without quantization.
\subsection{Fusion of Communicated Posteriors}
\label{subsec:fusion_posteriors}

At time step $t$, the communicated messages are fused into a common Gaussian estimate. Because all agents receive the same set of messages and apply the same fusion rule in the same order, they obtain the same fused result. Let
\begin{equation}
\mathcal{M}_{i,t} = (\bm{\mu}_{i,t}^{\mathrm{msg}}, \bm{P}_{i,t}^{\mathrm{msg}})
\label{eq:generic_message}
\end{equation}
denote the message of agent $i$, where
\begin{equation}
(\bm{\mu}_{i,t}^{\mathrm{msg}}, \bm{P}_{i,t}^{\mathrm{msg}})
=
\begin{cases}
(\bm{\mu}^{\,q}_{i,t|t}, \tilde{\bm{P}}_{i,t|t}), & \text{finite-rate case},\\[0.3em]
(\bm{\mu}_{i,t|t}, \bm{P}_{i,t|t}), & \text{CI-perfect case}.
\end{cases}
\label{eq:message_case_split}
\end{equation}
Since cross-correlations among local estimates are not explicitly tracked, fusion is performed using CI.

Consider two Gaussian estimates with means $\bm{\mu}_1,\bm{\mu}_2$ and covariances $\bm{P}_1,\bm{P}_2$. In information form, CI constructs the fused information matrix, covariance, and mean as
\begin{subequations}
\begin{align}
\bm{J}(\omega) &= \omega \bm{P}_1^{-1} + (1-\omega)\bm{P}_2^{-1}, \qquad \omega \in [0,1], \label{eq:ci_info}\\
\bm{P}(\omega) &= \bm{J}(\omega)^{-1}, \label{eq:ci_cov}\\
\bm{\mu}(\omega) &= \bm{P}(\omega)\!\left(\omega \bm{P}_1^{-1}\bm{\mu}_1 + (1-\omega)\bm{P}_2^{-1}\bm{\mu}_2\right). \label{eq:ci_mean}
\end{align}
\label{eq:ci_pairwise}
\end{subequations}
where $\bm{J}(\omega)$ denotes the fused information matrix.

The fusion weight is selected by minimizing the fused uncertainty according to
\begin{equation}
\omega^\star = \arg\min_{\omega \in \Omega_G} \log\det\!\bigl(\bm{P}(\omega)\bigr),
\label{eq:ci_weight}
\end{equation}
over the grid
\begin{equation}
\Omega_G = \left\{\frac{j}{G-1} : j = 0,\dots,G-1 \right\},
\label{eq:ci_grid}
\end{equation}
where $G$ denotes the number of grid points used in the search over $[0,1]$.

To combine all $N$ communicated messages, CI is applied sequentially in ascending agent-index order. The fusion is initialized with the first message,
\begin{equation}
(\bm{\mu}^{f}_{t}, \bm{P}^{f}_{t})
\leftarrow
(\bm{\mu}_{1,t}^{\mathrm{msg}}, \bm{P}_{1,t}^{\mathrm{msg}}),
\label{eq:fusion_init}
\end{equation}
and for $i = 2,\dots,N$ the current fused pair is updated as
\begin{equation}
(\bm{\mu}^{f}_{t}, \bm{P}^{f}_{t})
\leftarrow
\operatorname{CI}\!\left(
(\bm{\mu}^{f}_{t}, \bm{P}^{f}_{t}),
(\bm{\mu}_{i,t}^{\mathrm{msg}}, \bm{P}_{i,t}^{\mathrm{msg}})
\right).
\label{eq:fusion_seq}
\end{equation}

This yields the fused belief
\begin{equation}
\bm{x}_t \mid \mathcal{H}^{\mathrm{net}}_t \sim \mathcal{N}(\bm{\mu}^{f}_{t}, \bm{P}^{f}_{t}),
\label{eq:fused_belief}
\end{equation}
whose position components are used in the performance evaluation.
\section{VoC Metrics}
\label{sec:voc_metrics}

All performance metrics are evaluated on the two-dimensional position component of the state. For any Gaussian state estimate with mean $\bm{\mu}_t$ and covariance $\bm{P}_t$, we denote the associated position estimate and position covariance matrix by
\begin{equation}
\hat{\bm{p}}_t =
\begin{bmatrix}
\mu_t^{(p_x)} & \mu_t^{(p_y)}
\end{bmatrix}^{\top},
\qquad
\bm{P}_t^{\mathrm{pos}}
=
\begin{bmatrix}
[\bm{P}_t]_{11} & [\bm{P}_t]_{12}\\
[\bm{P}_t]_{21} & [\bm{P}_t]_{22}
\end{bmatrix},
\label{eq:position_projection}
\end{equation}
i.e., the first two entries of the mean vector and the upper-left $2\times2$ block of the covariance matrix.

\subsection{Distortion}

The true target position is denoted by
\begin{equation}
\bm{p}_t =
\begin{bmatrix}
p_x(t) & p_y(t)
\end{bmatrix}^{\top}.
\label{eq:true_position}
\end{equation}
The instantaneous tracking distortion is defined as the squared Euclidean position error
\begin{equation}
d_t = \|\hat{\bm{p}}_t - \bm{p}_t\|_2^2,
\label{eq:inst_distortion}
\end{equation}
and the time-averaged distortion over one trajectory is
\begin{equation}
D = \frac{1}{T+1}\sum_{t=0}^{T} d_t.
\label{eq:time_avg_distortion}
\end{equation}

We evaluate this quantity for the no-communication baseline of each agent, denoted by $D_{\mathrm{no},i}$, for the communication-assisted fused estimate at communication rate $R$, denoted by $D_{\mathrm{comm}}(R)$, and for the centralized oracle, denoted by $D_{\mathrm{oracle}}$. In addition, we report the CI-perfect reference $D_{\mathrm{CI\text{-}perf}}$, which corresponds to CI fusion without quantization under the same fusion architecture. This reference provides a quantization-free CI benchmark and helps distinguish quantization-induced loss from the residual loss due to conservative CI-based fusion.
\subsection{Gap to Oracle}

For any estimator with distortion $D$, we define the gap to the oracle as
\begin{equation}
\Delta = D - D_{\mathrm{oracle}}.
\label{eq:gap_oracle}
\end{equation}
Accordingly,
\begin{subequations}
\begin{align}
\Delta_{\mathrm{comm}}(R) &= D_{\mathrm{comm}}(R) - D_{\mathrm{oracle}}, \label{eq:gap_comm}\\
\Delta_{\mathrm{no},i} &= D_{\mathrm{no},i} - D_{\mathrm{oracle}}, \label{eq:gap_no}\\
\Delta_{\mathrm{CI\text{-}perf}} &= D_{\mathrm{CI\text{-}perf}} - D_{\mathrm{oracle}}. \label{eq:gap_ci_perf}
\end{align}
\label{eq:gap_cases}
\end{subequations}
Since $D$ is defined as a time-averaged squared position error, $\Delta$ has the same unit, i.e., $\mathrm{m}^2$ when position is measured in meters. It therefore quantifies the remaining estimation loss relative to the centralized oracle.
\subsection{Information Gain}

In addition to estimation error, we quantify the reduction in posterior uncertainty achieved by fusion. This is evaluated on the position covariance using a log-determinant criterion. For agent $i$, the per-step information gain from its local posterior to the communication-assisted fused posterior at rate $R$ is
\begin{equation}
I_{i\rightarrow\mathrm{comm}}(t;R)
=
\frac{1}{2}
\left[
\log\det\!\bigl(\bm P^{\mathrm{pos}}_{i,t|t}\bigr)
-
\log\det\!\bigl(\bm P^{\mathrm{pos}}_{\mathrm{comm},t}(R)\bigr)
\right],
\label{eq:info_gain_comm}
\end{equation}
and the corresponding local-to-oracle information gain is
\begin{equation}
I_{i\rightarrow\mathrm{oracle}}(t)
=
\frac{1}{2}
\left[
\log\det\!\bigl(\bm P^{\mathrm{pos}}_{i,t|t}\bigr)
-
\log\det\!\bigl(\bm P^{\mathrm{pos}}_{\mathrm{oracle},t}\bigr)
\right].
\label{eq:info_gain_oracle}
\end{equation}
Under the Gaussian entropy interpretation, the resulting values are measured in nats. For reporting, we average $I_{i\rightarrow\mathrm{comm}}(t;R)$ and $I_{i\rightarrow\mathrm{oracle}}(t)$ over $t=0,\dots,T$, yielding $\bar I_{i\rightarrow\mathrm{comm}}(R)$ and $\bar I_{i\rightarrow\mathrm{oracle}}$, respectively.
Positive values indicate that fusion reduces position uncertainty relative to the local estimate of agent $i$.
\subsection{Value of Communication per Bit}

The central VoC metric is the reduction in time-averaged position distortion achieved through communication, normalized by the communication rate. For agent $i$, it is defined for $R>0$ as
\begin{equation}
V_i(R)
=
\frac{D_{\mathrm{no},i} - D_{\mathrm{comm}}(R)}{R},
\label{eq:voc_per_bit}
\end{equation}
where $R$ denotes the per-agent communication rate in bits per time step. The quantity $V_i(R)$ therefore has unit $\mathrm{m}^2/\mathrm{bit}$ and measures the reduction in time-averaged squared position error relative to the no-communication baseline per transmitted bit.
\subsection{Additional Diagnostics and Monte Carlo Aggregation}

In addition to the core VoC measures, we report several auxiliary diagnostics. The root-mean-square error (RMSE) is defined as
\begin{equation}
\mathrm{RMSE} = \sqrt{D},
\label{eq:rmse_def}
\end{equation}
and therefore has unit $\mathrm{m}$. We also evaluate the empirical cumulative distribution function (CDF) of the instantaneous position error
\begin{equation}
e_t = \|\hat{\bm p}_t - \bm p_t\|_2,
\label{eq:inst_position_error}
\end{equation}
which provides a distributional view beyond the time-averaged squared-error metric.

To reflect sensing geometry, we report the position-level Fisher-information proxy
\begin{equation}
\bm J_{\mathrm{pos}}(t)
=
\sum_{i=1}^{N}
\bigl(\bm H^{\mathrm{pos}}_{i,t}\bigr)^{\top}
\bm R_i^{-1}
\bm H^{\mathrm{pos}}_{i,t},
\label{eq:fisher_proxy}
\end{equation}
where $\bm H^{\mathrm{pos}}_{i,t}$ is the position part of $\bm H_{i,t}$. We summarize it by $\log\det\!\bigl(\bm J_{\mathrm{pos}}(t)\bigr)$.

All metrics are aggregated over Monte Carlo runs and reported as sample means with approximate $95\%$ confidence-interval half-width
\begin{equation}
\mathrm{CI}_{95} = 1.96\,\frac{\hat{\sigma}}{\sqrt{n}},
\label{eq:ci95}
\end{equation}
where $\hat{\sigma}$ is the sample standard deviation and $n$ is the number of realizations.
\section{Results and Discussion}
\subsection{Experimental Setup}

We consider a 2D multi-agent tracking scenario in a $500\,\mathrm{m}\times500\,\mathrm{m}$ observation area. Three fixed agents observe a target through range and bearing measurements; they are intentionally clustered rather than uniformly spread, so that the joint sensing geometry varies strongly along the trajectory, favorable near the cluster and poorer as the target departs. The target follows the CV model introduced in Section~\ref{subsec:target_dynamics}, and performance is evaluated under finite-rate communication with quantized posterior messages and CI-based fusion. Unless stated otherwise, all simulation and evaluation parameters are summarized in Table~\ref{tab:exp_setup}. Performance is averaged over $100$ Monte Carlo runs and reported for the no-communication, oracle, CI-perfect, and finite-rate cases. The range and bearing noise levels are chosen to induce heterogeneous local tracking quality across agents. 

\begin{table}[!b]
\centering
\caption{Experimental setup and simulation parameters.}
\label{tab:exp_setup}
\begin{tabular}{ll}
\toprule
Parameter & Value \\
\midrule
Scene size & $500\times500\,\mathrm{m}^2$ \\
Sampling interval $\Delta t$ & $1.0$ $\mathrm{s}$\\
Horizon $T$ & $120$ $\mathrm{s}$ \\
Number of agents $N$ & $3$ \\
Initial target state $\bm x_0$ & $[0.0,\;0.0,\;1.2,\;0.7]^\top$ \\
Agent positions &
$\bm s_1=[-10,\,-80]^\top\,\mathrm{m}$ \\
& $\bm s_2=[170,\,0]^\top\,\mathrm{m}$ \\
& $\bm s_3=[10,\,80]^\top\,\mathrm{m}$ \\
Process noise std. $\sigma_a$ & $0.4$ $\mathrm{m/s^2}$ \\
Range noise std. & $[6.0,\;8.0,\;12.0]\,\mathrm{m}$ \\
Bearing noise std. & $[0.06,\;0.07,\;0.09]\,\mathrm{rad}$ \\
Rate per step per agent  $R=4b$ & $\{4,8,12,16,24,32\}$ bits  \\
Quantizer clipping factor $k_\sigma$ & $4.0$ \\
CI weight grid size & $101$ \\
\bottomrule
\end{tabular}
\end{table}
\subsection{Scenario Overview}

\begin{figure}[!t]
\centering
\subfloat[Agent 1 overview\label{fig:world_a0}]{
    \includegraphics[width=0.47\columnwidth,trim={0.cm 0.cm 0.cm 0.cm},clip]{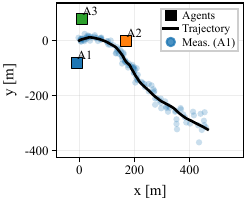}
}
\hfill
\subfloat[Agent 2 overview\label{fig:world_a1}]{
    \includegraphics[width=0.47\columnwidth,trim={0cm 0.cm 0.0cm 0cm},clip]{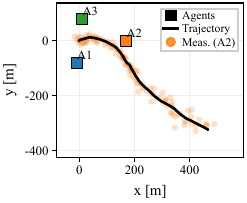}
}

\vspace{0.4em}

\subfloat[Agent 3 overview\label{fig:world_a2}]{
    \includegraphics[width=0.47\columnwidth,trim={0cm 0cm 0.cm 0.cm},clip]{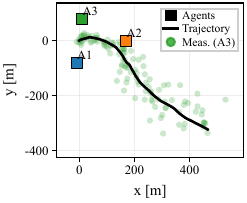}
}
\hfill
\subfloat[Geometry proxy\label{fig:geom_time}]{
    \includegraphics[width=0.47\columnwidth,trim={0.cm 0.cm 0.cm 0.cm},clip]{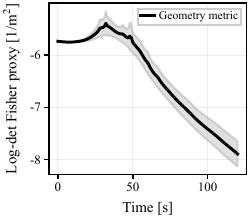}
}
\caption{(a)--(c) Scenario overview for the three agents, showing the target trajectory, fixed agent locations, and corresponding measurement clouds. (d) Geometry proxy over time, indicating that the joint sensing configuration becomes progressively less informative as the target moves through the scene.}
\label{fig:world_geom}
\end{figure}

Fig.~\ref{fig:world_geom}(a)--(c) illustrates the considered tracking scenario from the perspective of the three fixed agents. The target follows a curved trajectory through the scene, while the agent-wise measurement clouds reflect heterogeneous sensing conditions across agents and over time. In particular, Agents~1 and~2 exhibit comparatively smaller measurement spread over large parts of the trajectory, whereas Agent~3 shows larger dispersion, especially in the later part of the track.

Fig.~\ref{fig:world_geom}(d) summarizes the evolving joint sensing quality through the geometry-aware Fisher-information proxy $\log\det(\bm J_{\mathrm{pos}}(t))$. After a short initial increase, the metric decreases steadily, indicating that the aggregate sensing configuration becomes progressively less informative as the target moves through the scene. This non-uniform geometry is central to the subsequent VoC analysis: when sensing conditions are favorable, local estimates are already informative and the additional benefit of communication remains limited, whereas under less informative geometry, communication becomes more valuable.

\subsection{RMSE Performance and Rate Tradeoff}

\begin{figure}[!t]
\centering
\begin{minipage}{0.48\columnwidth}
    \centering
    \subfloat[RMSE over time\label{fig:rmse_timeseries}]{
        \includegraphics[width=\linewidth]{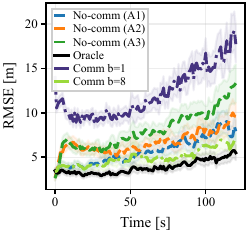}
    }
\end{minipage}
\hfill
\begin{minipage}{0.48\columnwidth}
    \centering
    \subfloat[RMSE vs.\ rate\label{fig:rmse_rate}]{
        \includegraphics[width=\linewidth]{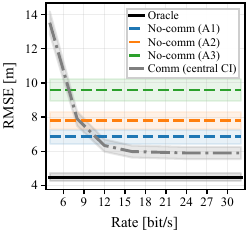}
    }
\end{minipage}
\caption{RMSE results: (a) temporal evolution for local baselines, oracle, and representative communication rates; (b) time-averaged RMSE versus communication rate.}
\label{fig:rmse_results}
\end{figure}

Fig.~\ref{fig:rmse_results}(a) shows the RMSE evolution over time for the local no-communication baselines, the centralized oracle, and the communication-assisted fused estimator at different quantization rates. The oracle remains best throughout the trajectory, while the three no-communication curves reveal clear agent heterogeneity: Agent~1 provides the best local baseline, followed by Agent~2, whereas Agent~3 exhibits the weakest local performance. In addition, the RMSE of all estimators increases toward the later part of the trajectory, consistent with the deteriorating sensing geometry in Fig.~\ref{fig:geom_time}.

The effect of communication depends strongly on the rate. At the lowest rate (\(b=1\), corresponding to \(R=4\,\text{bit/s}\) per agent), the communication-assisted estimator provides little benefit and is in fact outperformed by the local baselines over the trajectory. Table~\ref{tab:rate_perf} confirms that this regime is clearly harmful, with an RMSE of $13.93\pm0.72$~m and a very large gap to the oracle. Increasing the rate to \(R=8\,\text{bit/s}\) per agent already yields a major improvement, but the communication-assisted estimate still lies in a transitional regime. Here, increasing $R$ changes only the number of bits used to represent each communicated state component; the sensing interval, local EKF update frequency, and trajectory length remain fixed across all rate settings.

Fig.~\ref{fig:rmse_results}(b) together with Table~\ref{tab:rate_perf} shows that most of the performance gain is achieved in the low-to-moderate rate regime. At \(R=12\,\text{bit/s}\) per agent, the RMSE drops to $6.35\pm0.43$~m, marking a clear knee region in which communication already outperforms the local no-communication baselines. Beyond this point, the curve flattens: increasing the rate from $16$ to $32$ improves the RMSE only marginally, from $6.00\pm0.42$~m to $5.93\pm0.42$~m. Thus, moderate communication already captures most of the achievable gain, whereas additional rate yields only limited further benefit.

\begin{table}[t]
\centering
\caption{Communication-assisted performance versus rate (mean $\pm$ 95\% CI).}
\label{tab:rate_perf}
\begin{tabular}{cccc}
\toprule
\(R\,[\mathrm{bit/s}]\) per agent & RMSE [$\mathrm{m}$] & $\Delta_{\mathrm{comm}}$ [$\mathrm{m}^2$] & Regime \\
\midrule
4  & $13.93 \pm 0.72$ & $172.54 \pm 18.51$ & harmful \\
8  & $8.00 \pm 0.44$  & $45.41 \pm 5.77$   & transitional \\
12 & $6.35 \pm 0.43$  & $22.13 \pm 4.36$   & knee \\
16 & $6.00 \pm 0.42$  & $17.59 \pm 4.13$   & near saturation \\
24 & $5.94 \pm 0.42$  & $16.66 \pm 4.20$   & fusion-limited \\
32 & $5.93 \pm 0.42$  & $16.66 \pm 4.22$   & fusion-limited \\
\bottomrule
\end{tabular}
\end{table}
\subsection{Oracle Gap, Information Gain, and Communication Efficiency}

\begin{figure}[!t]
\centering
\subfloat[Gap to oracle vs.\ rate\label{fig:gap_rate}]{
    \includegraphics[width=0.47\columnwidth]{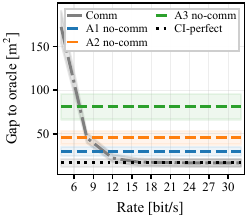}
}
\hfill
\subfloat[Information gain vs.\ rate\label{fig:ig_rate}]{
    \includegraphics[width=0.47\columnwidth]{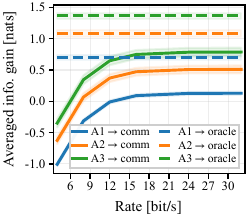}
}

\vspace{0.4em}

\subfloat[CDF of position error\label{fig:cdf_error}]{
    \includegraphics[width=0.47\columnwidth]{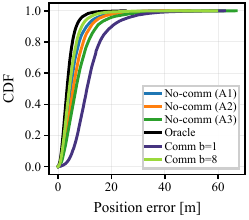}
}
\hfill
\subfloat[VoC per bit vs.\ rate\label{fig:voc_rate}]{
    \includegraphics[width=0.47\columnwidth]{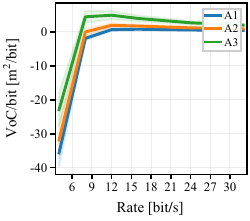}
}
\caption{Rate-dependent communication diagnostics: (a) gap to oracle, (b) information gain, (c) CDF of position error, and (d) VoC per bit.}
\label{fig:comm_diag}
\end{figure}

Fig.~\ref{fig:comm_diag}(a) shows that the communication-assisted fused estimator is highly rate-sensitive. At the lowest rate (\(R=4\,\text{bit/s}\) per agent), the oracle gap is largest, and the communication-assisted estimate even performs worse than the no-communication baselines. Table~\ref{tab:rate_perf} confirms that this low-rate regime is clearly harmful. As the rate increases, the oracle gap drops sharply, and by about $R=12$--$16$ the communication-assisted curve approaches the CI-perfect reference. Overall, these results suggest that most quantization loss is removed already at moderate rates, after which the residual gap is largely attributable to the fixed CI fusion architecture.

This interpretation is consistent with the information-gain results in Fig.~\ref{fig:comm_diag}(b). At \(R=4\,\text{bit/s}\) per agent, the local-to-communication information gain is negative for all agents, meaning that the fused estimate is less informative than the corresponding local posterior. As the rate increases, the information gain becomes positive and then gradually saturates. Table~\ref{tab:info_voc} further shows that this effect is strongest for Agent~3 and weakest for Agent~1, indicating that the locally weakest agent benefits most from communicated information.

A distributional view is provided by Fig.~\ref{fig:comm_diag}(c), where left-shifted curves indicate smaller position errors. The oracle dominates all other estimators, while the $R=4$ communication curve is clearly worst, confirming that very low-rate communication degrades performance not only on average but across most quantiles. In contrast, the $R=32$ communication curve is substantially left-shifted relative to the no-communication baselines and approaches the oracle much more closely, in agreement with the RMSE and oracle-gap results.

Finally, Fig.~\ref{fig:comm_diag}(d) with Table~\ref{tab:info_voc} highlights the communication-efficient region. The fall of error with rate until saturation is expected. The results reveal three regimes: a harmful low-rate regime in which communication drops below the no-communication baselines through the quantization-fusion interaction; an efficient intermediate regime marked by the peak of the non-monotone VoC-per-bit; and a high-rate regime of diminishing returns whose residual gap is fusion-limited rather than rate-limited, indicating whether remaining error is better addressed by more rate or better fusion.

\begin{table}[t]
\centering
\caption{Local-to-communication information gain and value of communication per bit versus rate.}
\label{tab:info_voc}
\setlength{\tabcolsep}{3.5pt}
\footnotesize
\begin{tabular}{ccccccc}
\toprule
\multirow{2}{*}{\(R\) [bit/s] per agent} & \multicolumn{3}{c}{Information gain [nats]} & \multicolumn{3}{c}{VoC per bit [$\mathrm{m}^2$/bit]} \\
\cmidrule(lr){2-4}\cmidrule(lr){5-7}
 & agent 1 & agent 2 & agent 3 & agent 1 & agent 2 & agent 3 \\
\midrule
4  & $-1.01$ & $-0.63$ & $-0.35$ & $-35.62$ & $-31.82$ & $-22.90$ \\
8  & $-0.31$ & $0.07$  & $0.34$  & $-1.92$  & $-0.02$  & $4.44$ \\
12 & $-0.01$ & $0.37$  & $0.65$  & $0.66$   & $1.93$   & $4.90$ \\
16 & $0.09$  & $0.47$  & $0.75$  & $0.78$   & $1.73$   & $3.96$ \\
24 & $0.13$  & $0.51$  & $0.79$  & $0.56$   & $1.19$   & $2.68$ \\
32 & $0.13$  & $0.51$  & $0.79$  & $0.42$   & $0.89$   & $2.01$ \\
\bottomrule
\end{tabular}
\end{table}
\subsection{Scope and Limitations}

The stationary agents and smooth target deliberately isolate rate and geometry, making these results a controlled baseline rather than a vehicular evaluation; the operating points do not transfer directly to mobile settings. Less conservative fusion, adaptive quantization, and a non-ideal channel would respectively lower the residual gap, reduce low-rate loss, and reduce the realized gain, so the measured VoC is best read as an upper reference. Still, the framework targets a question central to vehicle-to-everything and vehicle-to-vehicle (V2X/V2V) communication: how much cooperative communication a shared radio budget justifies.
\section{Conclusion}

This paper characterized the value of communication (VoC) in cooperative multi-agent tracking under rate-constrained belief exchange, using local extended
Kalman filtering (EKF), posterior quantization, and fixed
covariance intersection (CI) based fusion so that the measured behavior reflects the rate itself rather than a co-designed estimator. Against no-communication, oracle, and CI-perfect references, the results reveal three regimes: a harmful low-rate regime, an efficient intermediate regime with a clear knee near $R=12$--$16\,\text{bit/s}$ per agent, and a diminishing-return regime whose residual gap is fusion-limited rather than rate-limited. The benefit of communication also depends strongly on sensing geometry. These findings support the central message: communication is valuable for cooperative tracking only up to the point at which the achieved gain justifies its cost. Future work should pursue adaptive communication policies and mobility- and channel-aware models beyond the controlled setting  studied here.
\section*{Acknowledgment}
This work has been supported by the Federal Ministry of Research, Technology and Space of the Federal Republic
of Germany as part of the project Open6GHub+ under
contract 16KIS2404, and 6G-Sensoria. The work contributes
to the research within the 6G-Valley innovation cluster. The
authors alone are responsible for the content of the paper. The
authors used OpenAI ChatGPT to check grammar, spelling,
and improve readability and language.
\balance

\end{document}